\documentclass[prx,twocolumn,aps,floatfix,longbibliography,superscriptaddress,nofootinbib]{revtex4-1}
\usepackage[utf8]{inputenc}
\usepackage{natbib}
\usepackage[normalem]{ulem}
\usepackage{graphicx}
\usepackage{xcolor}

\usepackage[tbtags]{amsmath}
\usepackage{hyperref}
\hypersetup{colorlinks,allcolors=black}
\usepackage{amssymb}
\usepackage{gensymb}
\usepackage{float}
\usepackage{amsmath}
\usepackage{tabularx,graphicx}
\usepackage{epstopdf}
\usepackage{latexsym}
\usepackage{color, colortbl}
\usepackage{psfrag}
\usepackage{bbm}
\usepackage{bm}
\usepackage{titlesec}
\usepackage{dsfont}
\usepackage{feynmp}
\usepackage{slashed}
\usepackage{multirow}

\newcommand{\be}{\begin{equation}}
\newcommand{\ee}{\end{equation}}
\newcommand{\beq}{\begin{eqnarray}}
\newcommand{\eeq}{\end{eqnarray}}
\newcommand{\ba}{\[\begin{aligned}}
\newcommand{\ea}{\end{aligned}\]}

\newcommand{\s}{\sigma}
\newcommand{\la}{\langle}
\newcommand{\ra}{\rangle}

\newcommand{\V}{{\cal V}}

\renewcommand{\vec}[1]{{\bf #1}}
\renewcommand{\hat}[1]{{\bf {\widehat #1}}}
\renewcommand{\phi}{\varphi}
\renewcommand{\epsilon}{\varepsilon}

\def\nn{\nonumber}

\renewcommand{\vec}[1]{\boldsymbol{#1}}

\def \W{{\Omega}}

\def \g{{\gamma}}

\def \w{{\omega}}
\def \s{{\sigma}}

\def \r {\vec{r}}

\def \ra{{\rangle}}
\def \la{{\langle}}
\def \tn{\textnormal}

\def \ba{\begin{align*}}
\def \ea{\end{align*}}

\newcounter{indice}

\def \mrm{\mathrm}

\def \bs{\boldsymbol}
\def \mc{\mathcal}

\begin{document}
\title{Does filling-dependent band renormalization aid pairing in twisted bilayer graphene?}
\author{Cyprian Lewandowski}
\email[]{cyprian@caltech.edu}
\affiliation{Department of Physics, California Institute of Technology, Pasadena CA 91125, USA}
\affiliation{Institute for Quantum Information and Matter, California Institute of Technology, Pasadena CA 91125, USA}

\author{Stevan Nadj-Perge}
\affiliation{Institute for Quantum Information and Matter, California Institute of Technology, Pasadena  CA 91125, USA}
\affiliation{T. J. Watson Laboratory of Applied Physics, California Institute of Technology, 1200 East California Boulevard, Pasadena CA 91125, USA}

\author{Debanjan Chowdhury}
\email[]{debanjanchowdhury@cornell.edu}
\affiliation{Department of Physics, Cornell University, Ithaca NY 14853, USA}

\begin{abstract}
Magic-angle twisted bilayer graphene exhibits a panoply of many-body phenomena that are intimately tied to the appearance of narrow and well-isolated electronic bands. The microscopic ingredients that are responsible for the complex experimental phenomenology include electron-electron (phonon) interactions and non-trivial Bloch wavefunctions associated with the narrow bands. Inspired by recent experiments, we focus on two independent quantities that are considerably modified by Coulomb interaction driven band renormalization, namely the density of states and the minimal spatial extent associated with the Wannier functions. First, we show that a filling-dependent enhancement of the density of states, caused by band flattening, in combination with phonon-mediated attraction due to electron-phonon umklapp processes, increases the tendency towards superconducting pairing in a range of angles around magic-angle. Second, we demonstrate that the minimal spatial extent associated with the Wannier functions, which contributes towards increasing the superconducting phase stiffness, also develops a non-trivial enhancement due to the interaction induced renormalization of the Bloch wavefunctions. While our modeling of superconductivity assumes a weak electron-phonon coupling and does not consider many of the likely relevant correlation effects, it explains simply the experimentally observed robustness of superconductivity in the wide range of angles that occurs in the relevant range of fillings.
\end{abstract}

\maketitle


{\it Introduction.--} The origin of electron pairing in twisted bilayer graphene (TBG) has been
at the heart of the discussion since the discovery of superconductivity \cite{cao2,Yankowitz1059,efetov2019} in magic-angle twisted bilayer graphene (MATBG), while the phase diagram and the associated experimental phenomenology has continued to evolve  dramatically \cite{Balents2020}. At the time of writing of this letter, the following observations related to superconductivity (SC) are universally accepted in MATBG: (i) multiple pockets of SC are present over an extended range of fillings, $-4<\nu<4$ ($\nu\equiv$ electron filling in the moir\'e `flat' bands). The location of these SC regions are not simply tied to either the near vicinity of the correlation-induced insulators at commensurate fillings \cite{cao1,Yankowitz1059,efetov2019}, or to the van-Hove singularities (vHs) associated with the non-interacting bandstructure. (Note that although different models for the  non-interacting bandstructures may predict vHs to occur at different fillings, no single model predicts multiple vHs to be present at fillings corresponding to locations of the SC pockets.) 
(ii) The SC regions are more resilient to external screening and deviations away from magic-angle \cite{10.1038/s41586-020-2473-8,saito2019decoupling,stepanov2019interplay,liu2020tuning}, i.e. even when the sharp insulating gaps in the limit of low temperatures are no longer observable at the various commensurate fillings, SC continues to remain robust with only minor changes to the transition temperatures, $T_c$. 

\begin{figure*}[]
    \centering
    \includegraphics[width=\linewidth]{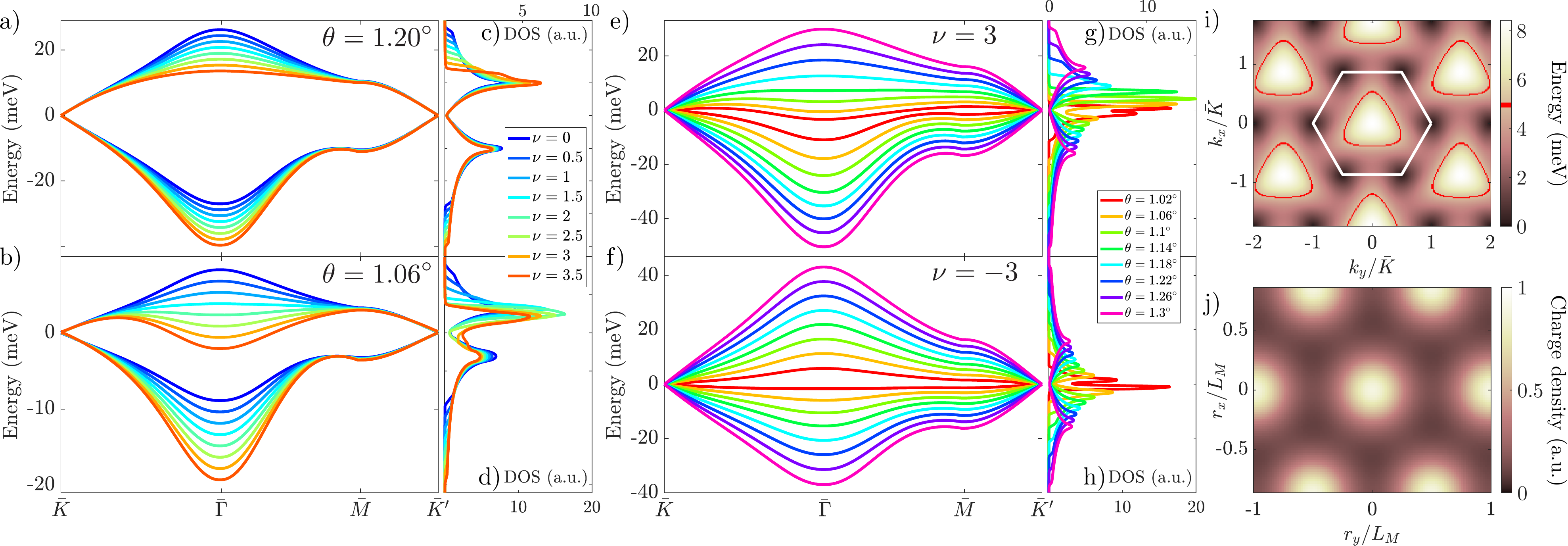}
    \caption{{\bf Hartree-Fock renormalized bandstructures.} TBG bandstructures as a function of filling for (a) $\theta =1.2^\circ$, and (b) $\theta = 1.06^\circ$ after including the HF corrections. One of the important features is related to band flattening and eventual inversion at the $\bar{\Gamma}$ point of the MBZ. The energy dependence of the density of states (c,d,g,h), demonstrating maximal enhancement when the band gets flattened, is shown next to each bandstructure. TBG bandstructures including HF corrections as a function of twist angle for (e) $\nu=3$, and (f) $\nu=-3$, respectively. In (a,b,e,f) we plot bandstructures for the $\xi = -1$ valley. i) The non-interacting energy landscape in the extended zone scheme of an electron band for
$\xi = -1$ at $\theta=1.06^\circ$. One MBZ is shown as a white hexagon. With red energy contour we denote Fermi surface for $\nu=3$. j) Charge density for a non-interacting bandstructure corresponding to $E_F = 4.92$ meV corresponding to the red energy contour in (i) and a filling of $\nu\approx 3$. Note that even at this large filling majority of charge density is located on the $AA$ sites of the effective triangular lattice giving rise to the large Hartree potentials. Here charge density is normalized by the highest charge density at full filling.}
\label{fig:intro}
\end{figure*}

Inspired by these experimental facts, we focus here on the following interesting theoretical scenario, where the sole effect of the electron-electron (Coulomb) interaction is to renormalize the bare non-interacting bandstructure in a filling-dependent fashion (see Fig.~\ref{fig:intro}a, b), while the attraction required for pairing stems from electron-phonon interactions. We capture the effects of these renormalizations on both the bandstructure and Bloch wavefunctions at the level of a Hartree-Fock (HF) approximation. As explained below, we match various qualitative aspects of the filling-dependent, renormalized bandstructure to recent experimental observations \cite{NadjPerge}. Within this setup, we will demonstrate that the superconducting phase-diagram as a function of density and twist-angle is markedly different from the one derived from a model of non-interacting bandstructure. With this plan in place, we are led to a number of important questions, that we address in this letter: (i) What controls the propensity towards pairing at angles away from magic-angle? (ii) To what extent does the bare electronic bandstructure influence the SC phase diagram and its pairing tendencies? (iii) What are the modified properties of the Bloch functions associated with the renormalized Hamiltonian and their possible effect on superconducting properties?

A number of recent theoretical works have focused on the role of bandstructure renormalizations in MATBG on the possible symmetry-broken insulating phases at commensurate fillings at the level of a HF approximation. However, the role of these renormalizations and especially the band ``flattening" behavior (to be made precise below; see Fig.~\ref{fig:intro}a-d) on the pairing tendencies has not been analyzed explicitly. In MATBG, as a result of the uneven real-space charge distribution within the unit cell that reflects the effective triangular symmetry of the TBG lattice (Fig.~\ref{fig:intro}e, f), the Hartree corrections become prominent \cite{guineaElectrostaticEffectsBand2018,rademakerChargeSmootheningBand2019,
ceaElectronicBandStructure2019,goodwinHartreeTheoryCalculations2020,
calderonInteractions8orbitalModel2020}. The exchange effects, which can lead to gap openings and change the topological properties, are directly accessible in transport experiments (e.g. Landau fan) \cite{Saito2021,Choi2021,das2020symmetry,wu2020chern,Nuckolls2020}. On the other hand, the Hartree corrections only alter qualitative aspects of the bandstructure and the underlying Bloch wavefunctions, leaving the topological properties unaltered and thereby making them harder to detect in transport. Interestingly, these changes can be imaged directly in local-probe experiments \cite{NadjPerge}. Here we build on this recently seen mechanism relying on the Hartree-correction, 
that modify the bandstructure and Bloch wavefunctions, and analyze their role on enhancing the tendency towards pairing in MATBG.

We note at the outset that we intentionally do not include the effects associated with the ``cascade transitions'' at integer fillings near magic-angle \cite{Ilani19,Yazdani19}, which
though relevant for concrete experimental observations will inevitably complicate further the discussion of superconductivity. As already indicated above, it is at present unclear to what extent the cascade is tied to the origin of SC; nevertheless, in what follows the doping dependence of SC near magic-angle will be modified in the vicinity of integer fillings where a cascade would be expected to occur. On the other hand, away from magic-angle where the effects of the cascade become less pronounced, the results for SC are less likely to change qualitatively (even though cascade-related signatures were observed down to low twist angles \cite{10.1038/s41586-020-2473-8}). For simplicity, we leave a careful analysis of the SC phase-diagram, including the effects of the cascade transitions, to a future study.  

{\it Renormalized bands.--} 
We begin with the single-particle continuum Hamiltonian \cite{MK18}: \beq
\mathcal{H}_0 &=& \sum_{\gamma = \{\xi,\sigma\}} \int_\Omega d^2\r~ \psi_\gamma^\dagger(\r)  \hat{H}^{(\xi,\sigma)} \psi_{\gamma}(\r),\\
\hat{H}^{(\xi,\sigma)} &=& \begin{pmatrix} 
H_{\xi 1} & U_{\xi}^\dagger(\r) \\
U_{\xi}(\r) & H_{\xi 2} 
\end{pmatrix},\label{eq:ham_cont}
\eeq
where the explicit form of $\hat{H}^{(\xi,\sigma)}$ appears in the Methods section. The spinor, $\psi_\gamma$, is written in the basis of $(A_1, B_1, A_2, B_2)$ sites of the original two layers ($l=1,2$) and we use the shorthand notation, $\gamma\equiv\{\xi(=\pm1),\sigma(=\pm1)\}$, for the valley/spin degrees of freedom. The real space integration is over a moir\'e unit cell $\Omega$. In what follows, any reference to the ``non-interacting model'' corresponds to a calculation that is based solely on the eigenvalues and eigenstates of this Hamiltonian in Eq.~\eqref{eq:ham_cont}.

The Coulomb interaction is given by, 
\beq\label{eq:ham_int}
\mathcal{H}_c &=& \frac{1}{2} \int_\Omega d^2\r~d^2\r'~\delta\rho(\r)~\V_{\mrm{c}}(\r-\r')~\delta\rho(\r'),\\
\delta\rho(\r) &=& \sum_\gamma \psi_\gamma^\dagger(\r)\psi_\gamma(\r) - \rho_{\tn{CN}}(\vec{r}).
\eeq
Here $\delta \rho(\vec{r})$ tracks the density relative to that at charge neutrality, $\rho_{\tn{CN}}(\vec{r})$, and $\V_{\mrm{c}}(\r-\r')$ is the Coulomb potential with a Fourier transform, $\V_{\mrm{c}}(\vec{q})=2\pi e^2 / \epsilon q$. For reasons to be made clear below, the dielectric screening by the substrate (denoted $\epsilon$) is treated as a free parameter.

We approximate the above interaction term using a self-consistent HF approximation as,
\beq
\mathcal{H}_c &\approx& \mathcal{H}_H + \mathcal{H}_F \equiv  \Sigma_{\tn{HF}}(\nu),
\label{Hc}
\eeq
where the many-body renormalization, $\Sigma_{\tn{HF}}(\nu)$, will lead to a modified electronic bandstructure due to either the Hartree ($\mathcal{H}_H$) or Fock ($\mathcal{H}_F$) terms, respectively. The Hartree correction is given by,
\beq
\mathcal{H}_H &=& \sum_\gamma \int_\Omega d^2\r ~V_H(\r)~\psi^\dagger_\gamma(\r)\psi_\gamma(\r),\\
V_H(\r) &=& \int_\Omega d^2\r'~\V_{\mrm{c}}(\r-\r')\sum_\gamma \left\langle \psi_\gamma^\dagger(\r')\psi_\gamma(\r')\right\rangle_{H}
\label{eq:Hartree}
\eeq
where $\la...\ra_H$ denotes a summation over occupied states measured from CNP ($\nu = 0$) \cite{guineaElectrostaticEffectsBand2018}.  
As a function of increasing doping relative to charge neutrality, there is a preferential buildup of charge at $AA$ sites in real space (Fig.~\ref{fig:intro}f), corresponding to electronic states near $\bar{K}$ points of the mini-Brillouin zone. The non-uniform spatial charge distribution generates an electrostatic potential that prefers an even redistribution of the electron density. In contrast, the real space charge distribution corresponding to electronic states near $\bar{\Gamma}$ point is more uniform in the unit cell. The effect of the electrostatic Hartree potential and the associated charge  redistribution thus leads to a lowering of the energy of the electronic states near the $\bar{\Gamma}$ point compared to the energy of states near the $\bar{K}$ points (Fig.~\ref{fig:intro}a, b).

The effect of the Hartree potential becomes increasingly pronounced as a function of decreasing twist-angle, especially near the magic-angle where the non-interacting bandwidth is minimal. There is an increasing tendency towards band-inversion near the $\bar{\Gamma}$ point \cite{ceaElectronicBandStructure2019,goodwinHartreeTheoryCalculations2020}, a feature that has not been observed in experiments till date \cite{NadjPerge}. However, it is important to note that the Fock term, $\mathcal{H}_F$, inherently acts against this tendency towards band-inversion via two key mechanisms \cite{xieWeakfieldHallResistivity2020}: (i) by increasing the overall bandwidth, and (ii) by contributing an opposing correction to the self-energy as compared to the Hartree term, Eq.~\eqref{eq:Hartree}.

The Fock term, $\mathcal{H}_F$, is given by,
\beq
\mathcal{H}_F = \sum_{\gamma}\sum_{i,j}\int_\Omega~d^2\r~d^2\r' V_{F,\gamma}^{ij}(\r,\r')~\psi_\gamma^{i\dagger}(\r) \psi_{\gamma}^j(\r'),\\
V_{F,\gamma}^{ij}(\r,\r') = -\V_{\mrm{c}}(\r-\r') \left\langle \psi_\gamma^{j\dagger}(\r')\psi^i_\gamma(\r)\right\rangle_F.
\label{eq:Fock}
\eeq
Note that the Fock potential, unlike Hartree, does not contain a summation over valley/spin degrees of freedom and as a result of the block-diagonal nature of the non-interacting Hamiltonian (Eq.~\eqref{eq:ham_cont}) does not contain any inter-flavour terms. We explicitly forbid any inter-flavour terms to be generated spontaneously \cite{ceaBandStructureInsulating2020,xieWeakfieldHallResistivity2020}, since our goal here is to focus on the qualitative changes to the band structure and not on determining the precise nature of the correlated insulators \cite{bultinckGroundStateHidden2020}. Finally the notation $\left\langle \dots \right\rangle_F$ corresponds to a summation over occupied states; see Methods for a more detailed discussion of the subtleties and the various conventions adopted in earlier works regarding the Fock term.

Our modeling of the bandstructure is motivated by recent experiments \cite{NadjPerge}; see Methods for details. In particular, we determine the microscopic parameters for the model by matching our theoretical bandstructures to the experimental results sufficiently far away from the magic-angle. These parameters are kept fixed for all twist angles and as a result we do not capture the subtle lattice-relaxation effects near magic-angle \cite{carrExactContinuumModel2019,fangAngleDependentItInitio2019}. Note also that as a result of our adopted procedure, the location of magic angle is $\theta\approx1^\circ$, which is different from the value encountered most often in literature, $\theta\approx 1.1^\circ$. At the same time, it is worth noting that in the continuum model, varying the ratio of the interlayer parameters does not drastically alter the location of van Hove singularities. It is thus possible, in principle, to disentangle the effects due to twist dependent interlayer hopping ratio from those due to band-flattening physics (see Supp. Mat. Sec. \ref{methods: parameters} for further discussion). For general agreement with the experimental results, we found it necessary to use a dielectric constant $\epsilon$ larger than that set by the substrate, in accordance with similar observations made in earlier studies \cite{xieWeakfieldHallResistivity2020,guineaElectrostaticEffectsBand2018,ceaElectronicBandStructure2019}.
In spite of these simplifying approximations, our modeling captures the qualitative behavior exhibited by the measured MATBG bandstructure as the twist angle is brought closer to the magic-angle condition (See Ref.\cite{NadjPerge} and discussion in Supp. Mat. Sec.\ref{methods: parameters}). The final renormalized bandstructures at fixed angles of $\theta=1.20^\circ,1.06^\circ$ are shown as a function of filling in Fig.~\ref{fig:intro}a, b. Similarly, for a fixed filling, the bandstructures for increasing twist angles are shown in Fig.~\ref{fig:intro}c, d. Most notably, we see that as the twist angle approaches magic-angle, the bandstructure becomes locally flat near the $\bar{\Gamma}$ point beyond a certain filling. At these fillings, the location of the vHs changes from that set by the non-interacting model parameters as a result of the HF renormalization, c.f. density of states panels in Fig.~\ref{fig:intro}. We note that Ref.~ \cite{klebl2020importance} has also studied the onset of band-flattening in MATBG and commented on its possible relevance for enhancing effects of interactions. For our ensuing discussion of phonon-mediated attraction, the non-trivial band flattening and associated interaction induced shift of the vHs will play a crucial role in determining the shape of the superconducting dome.

{\it Phonon-mediated attraction.--} 
With the role of electron-electron interactions limited only to the corrections discussed above, we now take these bandstructures and accompanying wavefunctions to investigate phonon-mediated pairing. Earlier works have highlighted the importance of a purely phonon driven mechanism, within various approximations, when the electronic bandstructure is limited to the non-interacting model 
\cite{TH18b,Fu18,TH18a,Wu18,HJC18,BAB19,PO20,CL20,PhysRevB.102.064501,TBGV}. 
We employ here the framework and notation of an earlier work by us, focusing exclusively on an intervalley, spin-singlet gap function with zero center of mass momentum for simplicity \cite{CL20}. The effective electron-electron interaction, after integrating out the phonons, has the form: 
\begin{align}\label{eq:ham_int_ph}
&\mc S_{\tn{int}} = \frac{1}{2}\sum_{\substack{\vec{q},\w\\ \xi,\xi'}}\V^{\mrm{ph}}_{\xi,\xi'}(\vec{q},i\w)\nn \rho_{\xi}(\vec{q},i\w)~\rho_{\xi'}(-\vec{q},-i\w),\\
&\rho_{\xi}(\vec{q},i\w) = \sum_{\bs k, \nu} \overline{\Lambda}_{\g\g'}(\bs k+\bs q,\bs k)~c^\dagger_{\nu+\w\vec{k}+\vec{q}\{\g\}} c_{\nu\vec{k}\{\g'\}}\,, 
\end{align}
where the projected density operators, $\rho_{\xi}(\vec{q},i\w)$, include the form-factors,
$\overline{\Lambda}_{\g\g'}(\bs p,\bs k)=\delta_{\xi\xi'}\delta_{\s\s'}\left\langle \bs p,\{\g\}\left|e^{i (\vec{p-k})\cdot \vec{r}} \right|\bs k,\{\g'\}\right \rangle$.  Here $i\omega$ is the fermionic Matsubara frequency and $|\bs k,\{\g'\} \rangle$ denotes a Bloch wavefunction of the mean-field Hamiltonian, which includes the HF renormalizations due to the Coulomb interactions.  
The phonon mediated interaction vertex is given by,
\be \label{eq:V_ph}
\V_{\xi,\xi'}^{\tn{ph}}(\vec{q},i\w) = -\tilde{g}\frac{\omega^2_{\tn{ph}}(q)}{ \w^2 + \omega^2_{\tn{ph}}(q)} \delta_{\xi,-\xi'},
\ee
where $\omega_{\tn{ph}}(q)=c_s q$ is the acoustic phonon dispersion for graphene.  
The electron-phonon coupling constant, $g = D^2/\rho_m c_s^2$, is related to the deformation potential, $D~(= 25~ \tn{eV})$, the speed of sound in graphene, $c_s~ (= 12 000~\tn{m/s})$ and atomic mass density, $\rho_m ~(= 7.6\times 10^{-8} ~\mathrm{g}/\mathrm{cm}^2)$  \cite{PhysRevLett.105.256805,Chen2008}. Here we have redefined $\tilde{g}=g/N$ with a large-$N$ prefactor to obtain a controlled theoretical limit in which the Eliashberg equations for the pairing gap function are asymptotically exact; see Ref.~\cite{CL20} for details of our earlier large-$N$ framework. In the large-$N$ formulation, the Hartree-Fock corrections of interest to us here are subleading in $1/N$ ~\cite{CL20}. On the other hand, partly inspired by the emerging phenomenological considerations pointing towards the importance of these corrections, we employ a two-step calculation procedure where the Hartree-Fock modifications to the bandstructure are included first at leading order in the interaction strength, followed by the second stage where the pairing computation is carried out using the Eliashberg approach ignoring vertex corrections.

The Eliashberg equation for the gap function, $\Delta(i\omega,{k})$, is given by
\begin{align}
 \Delta(i\omega,{k}) = -T \sum_{\omega'} \sum_{{p}} K(i\w,{k};i\omega',{p})\Delta(i\omega',{p}) \label{eq:sc_gap_equation},
\end{align}
with a kernel $K(...)$,
\begin{align}\label{eq:Kernel}
K(i\w,&{k};i\omega',{p}) \equiv \\&\frac{1}{(2\pi)^2} \int d\W_{\bs p} \V^{\tn{ph}}_{-\xi,\xi}(\bs k-\bs p,i\omega-i\omega') \frac{\overline{\Lambda}( \bs p ,\bs k)\overline\Lambda(-\bs p,\bs k) }{{\omega'}^2+{\overline{E}}^2_{\xi,\vec{p}}},\nn
\end{align}
where $\int d \W_{\bs p}$ denotes integration over the angle between vector $\vec{k}$ and $\vec{p}$ for a fixed direction of $\vec{k}$. Importantly, $\overline{E}_{\xi,\vec{p}}$ is the electronic dispersion including the density dependent HF renormalization. Note that it is important to include a summation over the moir\'e umklapp processes in Eq.\eqref{eq:Kernel}, that were shown to lead to an increase in the pairing scale as in Ref.\cite{CL20}. We summarize the role of the form-factors $\overline{\Lambda}(\dots)$ and the origin of this effect in the Supplementary materials and demonstrate its effect on pairing tendencies in the Supplementary Figure 3. 

\begin{figure}
    \centering
    \includegraphics[width=\linewidth]{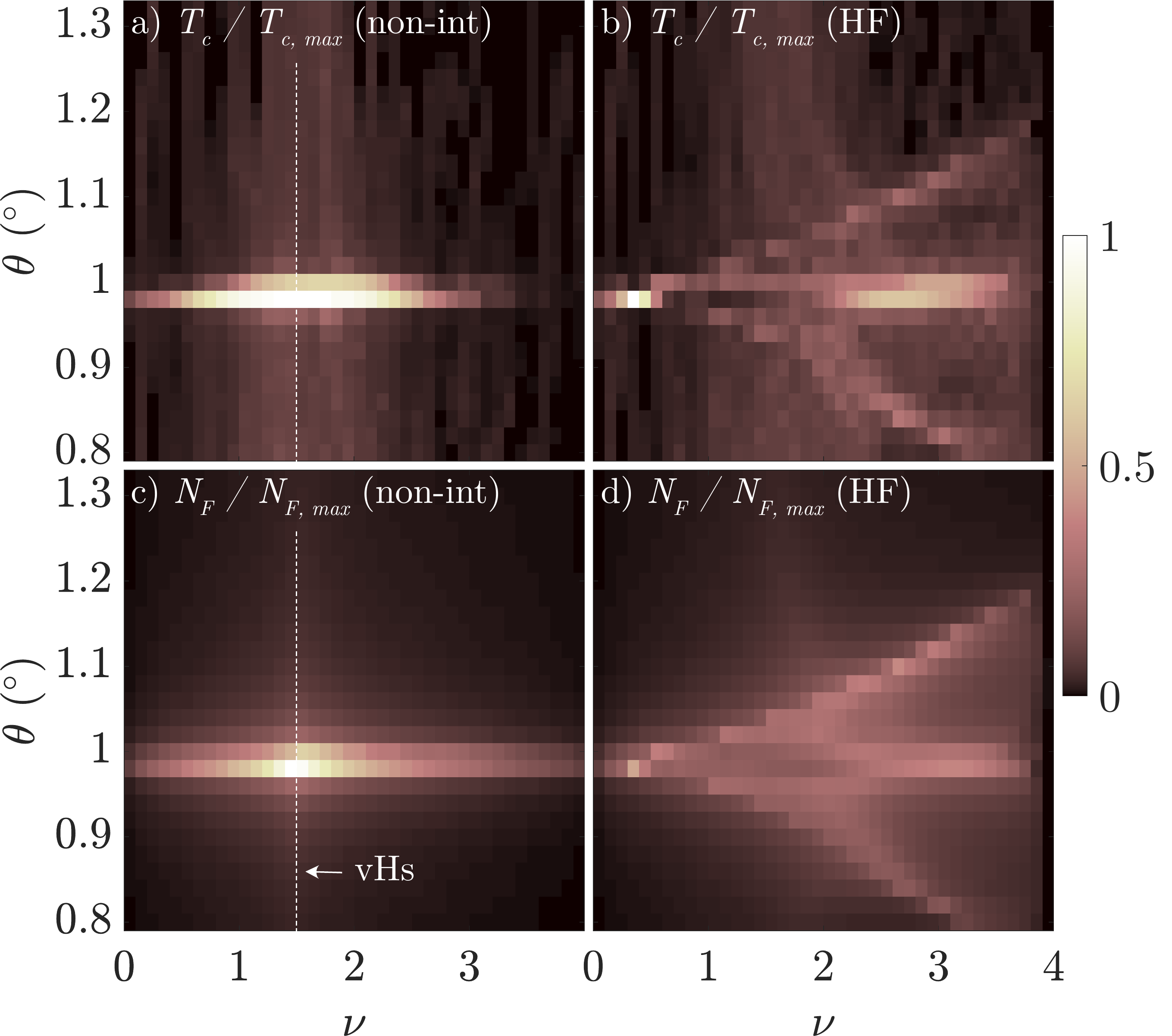}
    \caption{{\bf Pairing temperature with and without HF interactions.} $T_c$ obtained from the linearized Eliashberg equation (Eq.~\ref{eq:sc_gap_equation}) as a function of $\nu$ and $\theta$ for (a) non-interacting, and (b) HF-corrected bandstructure. The values of $T_c$ are normalized in both (a) and (b) relative to the highest pairing temperature $T_{C,max}$ in (a). Density of states at the Fermi level, $N_F$, as a function of $\nu$ and $\theta$ for (c) non-interacting, and (d) HF-corrected bandstructure. In (a) and (c), the vHs (labeled by white dashed line) is fixed by parameters of the non-interacting bandstructure and is peaked near the magic-angle. With band flattening, location of vHs in (d) becomes filling and twist-angle dependent. (See also Supplementary materials and Supplementary Figures 1, 2 for further comparison of the non-interacting and HF-corrected system)}. 
    \label{fig:results_T_C}
\end{figure}

{\it Robustness of SC away from magic-angle.--}
Within the framework of Eq.~\ref{eq:sc_gap_equation}-\ref{eq:Kernel}, $T_c$ is determined by the temperature at which the linearized Eliashberg equation (Eq.\ref{eq:sc_gap_equation}) has an eigenvalue of $1$. One of the central results of this paper appears in Fig.~\ref{fig:results_T_C}, which shows $T_c$ for (i) a non-interacting model without HF corrections (Fig.~\ref{fig:results_T_C}a), and (ii) a model that includes the density dependent HF corrections  (Fig.~\ref{fig:results_T_C}b), for a range of fillings and twist-angles. The results are qualitatively distinct; in particular, the intricate structure for $T_c$ as a function of $\theta$ and $\nu$ in Fig.~\ref{fig:results_T_C}b is seemingly unrelated to properties of the non-interacting bandstructure. For the latter, $T_c$ peaks at the magic-angle and then rapidly falls off with a varying twist angle. The peak of the SC dome as a function of $\nu$ is also pinned to be at the same filling. On the other hand, $T_c$ for the HF corrected bandstructure does not abruptly fall off with changing twist angle and its maximal value is not pinned at a fixed filling. The qualitative behavior for $\nu<0$, both with and without HF interactions, is similar to the corresponding results for $\nu>0$ (Fig.~\ref{fig:results_T_C}); the quantitative differences arise from the particle-hole asymmetry that is present between the electron and hole bands (see Supplementary materials for further details).

Within our Eliashberg analysis of the ``weak" electron-phonon coupling (controlled by large-$N$), $T_c$  is ultimately controlled by the density of states at the Fermi surface, $N(E_F)$, as shown in panels Fig.~\ref{fig:results_T_C} c, d for the same range of fillings and twist-angles.  
The interesting structure associated with $N(E_F)$ in Fig.~\ref{fig:results_T_C}d arises from the Hartree-contribution to the bandstructure. The phenomenon of band-flattening shifts the van Hove singularity in the density of states away from the location dictated by the parameters of the non-interacting band structure (bright feature indicated with arrows in Fig.~\ref{fig:results_T_C}c) and enhances the density of states at the Fermi level as shown in Fig.~\ref{fig:intro}a, b. As a function of decreasing twist angle (especially approaching magic-angle), the filling beyond which we see an onset of band flattening near $\bar{\Gamma}$ point (see Fig.~\ref{fig:intro}a, b) also decreases. This is evident from the ``three-prong" like feature in Fig.~\ref{fig:results_T_C}d. 
Remarkably, inclusion of the many-body renormalization to the bandstructure offers a plausible explanation for the robustness of SC over a broader range of twist angles, in line with experimental results\cite{cao2,Yankowitz1059,10.1038/s41586-020-2473-8,saito2019decoupling,stepanov2019interplay,Balents2020} and unlike the prediction of the bare unrenormalized bandstructures.

Before proceeding any further, we remind the reader that the apparent appearance of a two-peak-like structure at magic angle (Fig.~\ref{fig:results_T_C}b) will be masked by the onset of ``cascade-like" transitions occurring near integer fillings \cite{Ilani19,Yazdani19}; these latter effects have not been included in this study. In particular, this will lead to a suppression of band-flattening and, more crucially, induce a sequence of flavor selective cascade transitions that will alter the profile of density of states (and thereby $T_c$).  
Importantly, the Hartree-induced band-inversion near $\bar{\Gamma}$ becomes suppressed for $\nu > 1$ after a cascade transition. This in turn will likely  increase the density of states near $\nu \approx 2$ for twist angles near the magic-angle compared to the results of our current analysis (Fig.~\ref{fig:results_T_C}d), possibly resolving the apparent contradiction with the experimental results \cite{Balents2020}. We also anticipate that Cooper pairing will be unlikely for $\nu\gtrsim3$ due to the same underlying reasons. 

In our discussion thus far, we have identified $T_c$ with the onset of a pairing amplitude. In two-dimensions, the actual $T_c$ is determined by the superconducting phase stiffness. Thus even though our simplified analysis predicts an enhanced tendency towards SC near $\nu\sim0$ in the vicinity of magic-angle (Fig.~\ref{fig:results_T_C}b) and a peak that is shifting towards higher fillings at increasing twist angles ($\theta\gtrsim 1.15^\circ$), the SC phase stiffness will drop precipitously as the bands become nearly filled ($\nu\lesssim 4$) or approach charge neutrality ($\nu\sim0$). Nevertheless, we already see that band-flattening aids in the onset of the simplest phonon-induced attraction as a result of a renormalized density of states at the Fermi energy at larger twist angles.

\begin{figure}
    \centering
    \includegraphics[width=\linewidth]{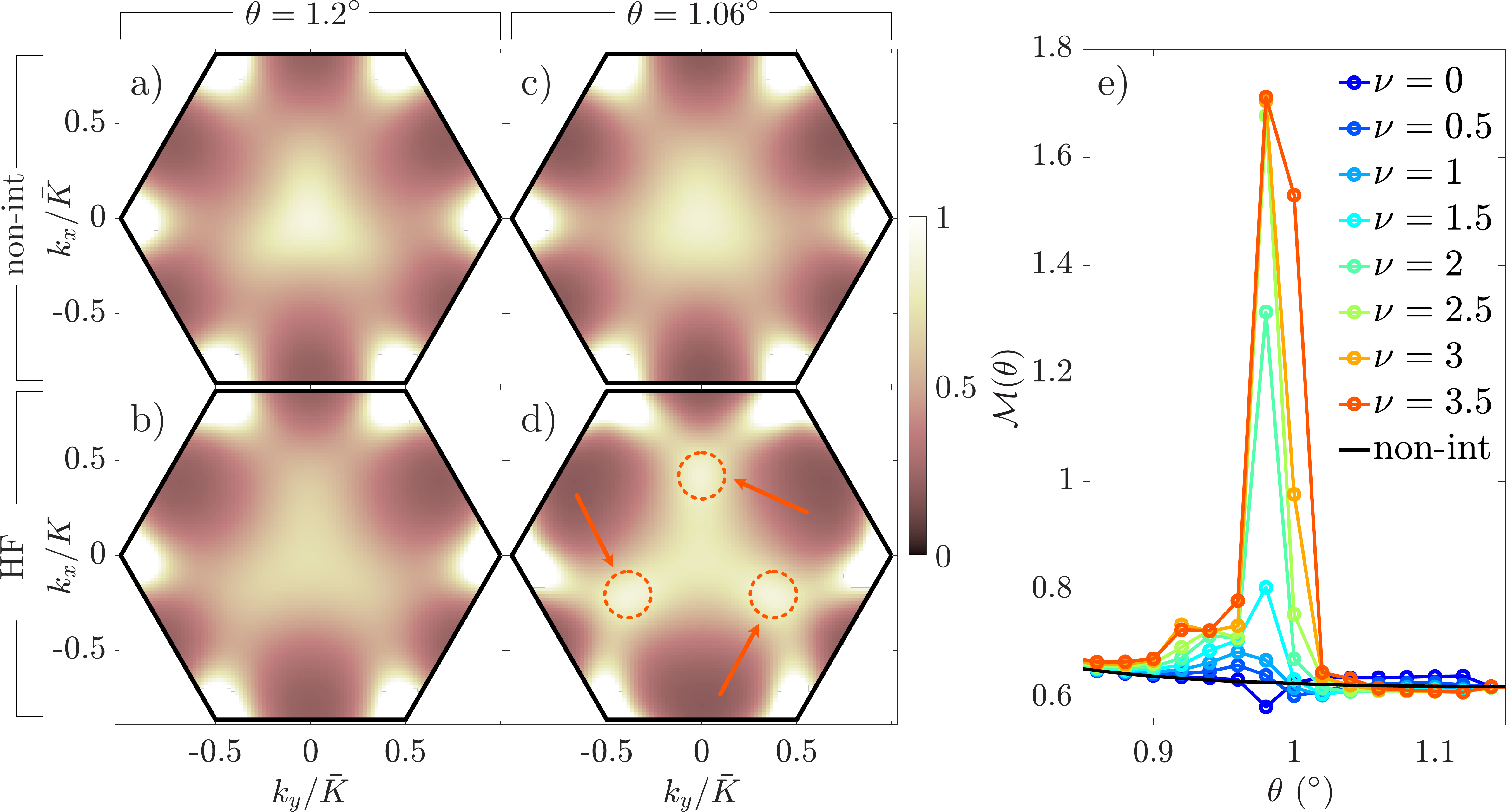}
    \caption{{\bf HF-driven modification of the Fubini-Study metric.} $\tn{Tr}(g_{\alpha\beta})$ for $\theta=1.20^\circ,1.06^\circ$ for (a,c) non-interacting, and (b,d) HF-corrected bandstructures for valley $\xi=-1$ at a filling of $\nu=4$. 
    The behavior changes qualitatively near magic-angle ($1.06^\circ$) with and without interactions. In panels (a)-(d), we saturate the color-scale to help delineate the features near the center of the MBZ. 
    e) A comparison of $\mathcal{M}$ for the HF-corrected bandstructure as a function of twist angle and filling with the non-interacting result. There is a dramatic enhancement in $\mathcal{M}$ for $\theta$ and $\nu$ where the tendency towards band inversion is most pronounced.}
    \label{fig:geometric}
\end{figure}

{\it Fubini-Study metric.--} 
As noted above, $T_c$ is ultimately determined by the superconducting phase stiffness. For large bandwidths, the higher electronic kinetic energy contributes to a larger phase stiffness. On the other hand, for narrower bands the problem becomes inherently non-perturbative and the exact mechanism that leads to a finite phase stiffness is presently unclear. However, if the Wannier functions are non-localizable \cite{Vanderbilt2} (e.g. as in topological bands), or have a finite geometric extent, the local Cooper pairs can contribute to the phase stiffness even in a perfectly flat-band limit \cite{DC20}. Within a BCS mean-field description of the projected interaction, the stiffness in the flat-band limit is proportional to the minimal spread of the Wannier functions \cite{Torma15,Bernevig19,Torma19,Rossi19}, which would also be the case for our weak-coupling computation. To investigate how the geometrical properties of the Bloch functions of the renormalized HF Hamiltonian evolve with density and twist-angle, we study the Fubini-Study (FS) metric,
\beq
\label{eq:fubini}
g_{\alpha\beta}(\vec{k}) &=& \frac{1}{2}\left(  u_{\vec{k},\alpha}^\dagger u_{\vec{k},\beta}+u_{\vec{k},\beta}^\dagger u_{\vec{k},\alpha}\right)+ u_{\vec{k}}^\dagger u_{\vec{k},\alpha} u_{\vec{k}}^\dagger u_{\vec{k},\beta}\,,\nonumber\\
&  \tn{where} & ~u_{\vec{k}}\equiv |\bs k,\{\g\} \rangle\,, u_{\vec{k},\alpha} \equiv \partial_{k_\alpha} u_{\vec{k}}\,,
\eeq
evaluated for the same flavor $\gamma$. The trace of the FS metric,   $\tn{Tr}(g_{\alpha\beta})$, controls the minimal spread of the associated Wannier functions via $\mathcal{M}=\int_{\vec{k}} \tn{Tr}(g_{\alpha\beta})$. The results are as shown in Fig.~\ref{fig:geometric}. 

The Bloch wavefunctions and the associated FS metric (Eq. \eqref{eq:fubini}) for the renormalized Hamiltonian, $\mathcal{H}_0+\mathcal{H}_c$, undergo a qualitative change from the corresponding quantities for the single-particle Hamiltonian, $\mathcal{H}_0$, as twist angle approaches magic-angle. Sufficiently far away from magic-angle (e.g. $\theta=1.20^\circ$), there is negligible difference between the metric for the non-interacting (Fig.~\ref{fig:geometric}a) vs. HF-modified bands (Fig.~\ref{fig:geometric}b). The subtle band-flattening features present in the bandstructures, Fig.~\ref{fig:intro}a, do not qualitatively alter the momentum dependence of the metric. Closer to the magic-angle (e.g. $\theta=1.06^\circ$) wherein the HF-modified bands develop band inversions (Fig.\ref{fig:intro}b), the modified Bloch wavefuctions lead to an appearance of distinct new features in the momentum dependence of the metric (see orange arrows and dashed circles in Fig.~\ref{fig:geometric}c, d). The locations of these new features correspond to the local maxima of the HF-corrected bandstructure. As a result, the integrated metric, ${\cal{M}}$, is also strongly modified in the vicinity of magic-angle (Fig.~\ref{fig:geometric}e) and might contribute towards an independent source of enhancement of the superconducting phase stiffness. Note that within the general weak-coupling setup, the contribution to the stiffness due to the enhanced spatial extent of Wannier-functions is independent of the underlying details of the pairing symmetry and phonon-mediated attraction.

It is important to note here that irrespective of whether the bands are topological or not, the minimal spatial extent of the Wannier functions is related to the integral of the trace of the FS metric. Regardless of the actual form, shape and other properties of the Wannier functions, the strong non-monotonic enhancement of this quantity as a function of twist angle near the magic-angle is an imporatant consequence of the band-flattening mechanism (Fig.\ref{fig:geometric}c).

We note that while the presence of cascade can partly obscure the large enhancement of the integrated metric near magic-angle, c.f. Fig.\ref{fig:geometric}e, we anticipate $\mathcal{M}$ to increase even when effects of cascade are taken into consideration. The enhancement of $\mathcal{M}$ is tied to incipient band-inversions near the $\bar{\Gamma}$ point, hints of which have been observed in recent experiments \cite{NadjPerge}. Interestingly, the distribution of the Berry curvature in the MBZ is also affected due to the HF corrections even though the valley projected Chern invariant remains unchanged; see Supplementary materials for further discussion. A recent experimental work \cite{pierce2021unconventional} highlights the role played by redistribution of Berry curvature in the MBZ as a key-element driving appearance of new insulating states.

{\it Discussion.--} 
In this work, we have focused on HF-driven corrections to the narrow electronic bands and their effect on phonon-induced pairing in the $s$-wave channel. However,  
phonons can also induce (weaker) attraction in other channels (e.g. $d$-wave) that will nevertheless exhibit qualitatively similar behavior as a function of filling and twist-angles (albeit with lower $T_c$), especially if the short distance part of the Coulomb repulsion suppresses $s$-wave pairing \cite{Wu18}.   
Beyond the weak-coupling limit considered here, it is difficult to speculate on the robustness of our conclusions and on the possibility of other competing instabilities in the absence of a non-perturbative analysis. 
Nonetheless, given the experimental evidence for filling-dependent band-flattening and our explicit demonstration of the dramatic many-body effects on the electronic bands and Bloch wavefunctions, it is clearly necessary to include them for a complete understanding of superconductivity in TBG, and moir\'e materials in general. (See Refs. \cite{2021arXiv210301815C,PhysRevMaterials.5.084008} for a more recent discussion of the role of band-flattening in other settings.)

Let us now comment on the subtle effects of an external gate-induced screening on the interplay between electron-electron interaction induced modifications to the bands and phonon-mediated SC for the same device without simultaneously varying twist-angle --- a question that was partly addressed in a recent experiment \cite{liu2020tuning}. Although the overall scale of both Hartree and Fock terms is controlled by the strength of the dielectric screening ($\epsilon$), the absolute magnitude of Hartree and Fock contributions can differ as they are dominated by scattering processes on different momentum scales. While the Hartree correction is controlled only by momenta on scales comparable to and larger than the moir{\'e} reciprocal lattice vector, $G_M=4\pi/\sqrt{3} L_M$, \cite{guineaElectrostaticEffectsBand2018,ceaElectronicBandStructure2019} ($L_M\equiv$ moir\'e period), the Fock term receives contribution from small (less than MBZ size) momentum processes as well. As such, when the screening gate is at a distance, $d \lesssim 1/2 G_M$, the Fock term will be more strongly suppressed relative to the Hartree term. Since the former is responsible for preventing band-flattening (and incipient band-inversion), we expect screening to enhance pairing tendencies. On the other hand, even if future experiments can resolve such changes in $T_c$, it might be difficult to disentangle these effects from a more conventional source of enhancement arising from suppression of the Coulomb repulsion.

It is worthwhile to comment on the regime of validity of some of the approximations we have made, specifically with regard to the temperature-dependent corrections in the Hartree-Fock analysis and the possible effects of including cascade-like transitions in the analysis here. The Hartree-Fock corrections are included at zero temperature, which is justified and self-consistent within our setup. Specifically, the leading temperature dependent corrections to the Hartree self-energy are small in $T/E_F$, where $E_F$ is defined as the chemical potential difference of the minority carriers measured with respect to the bottom/top of the band. At weak-coupling, the $T_c$ obtained by us is already much smaller than $E_F$, and including the additional corrections to the Hartree self-energy will not alter the results significantly. The origin of the Hartree corrections (and the associated band-flattening) is expected on more general grounds even at finite, but small, temperatures due to the inhomogenous charge distribution in real-space.

The effects of the cascade transitions, which have been neglected, can be included, in principle, in a future extension of the framework discussed here. The actual implementation would inevitably have to rely on a detailed input of the precise ground state near the integer fillings (see e.g. \cite{ceaBandStructureInsulating2020,bultinckGroundStateHidden2020,2021arXiv210401145K,PhysRevLett.122.246401})and the nature of the superconducting state in the vicinity of such a state. Our results for the FS metric are likely less sensitive to the cascades, since the resulting real space inhomogenous charge distribution and the associated band-flattening should continue to be relevant.

Finally, given the extent to which interactions can modify the Bloch wavefunctions and the associated quantum geometric tensor, it is desirable to design experiments suited towards moir\'e materials that can probe these quantities directly (regardless of whether, e.g. ${\cal{M}}$ determines the actual phase stiffness). In principle, transport or optical measurements can be used to infer the distribution of Berry curvature \cite{RevModPhys.82.1959,RevModPhys.82.1539,10.1038/nphys3551,10.1038/nature25987}. 
Similarly, it would be interesting to extract the FS metric by analyzing corrections to the predictions of semiclassical equations of motion in the presence of non-uniform electric fields away from magic-angle \cite{Hughes19}. Given that many of these techniques are not restricted to operate only at low temperatures,  
hopefully future experiments can extract these quantities and probe the physics of band-flattening well above the temperatures associated with cascade-transitions and superconductivity, 
thereby  
providing further direct insight into the complexities of twisted bilayer graphene. 

\acknowledgements
We thank Jason Alicea, Erez Berg, Johannes Hofmann and Pablo Jarillo-Herrero for useful discussions and Jonathan Ruhman for an earlier collaboration. C.L. acknowledges support from the Gordon and Betty Moore Foundation through Grant GBMF8682. S.N-P. acknowledges support from NSF (DMR-1753306), the Sloan Foundation, and the Institute for Quantum Information and Matter, an NSF Physics Frontiers Center with support of the Gordon and Betty Moore Foundation through Grant GBMF1250. 
D.C. is supported by a faculty startup grant at Cornell University.


\let\oldaddcontentsline\addcontentsline
\renewcommand{\addcontentsline}[3]{}
\bibliographystyle{naturemag}
\bibliography{references}
\let\addcontentsline\oldaddcontentsline

\clearpage
\onecolumngrid

\appendix

\renewcommand{\figurename}{Supplemental Figure}
\setcounter{figure}{0}
\begin{center}
    \bf{\Large Supplementary materials for ``Does filling-dependent band renormalization aid pairing in twisted bilayer graphene?"}
\end{center}

\tableofcontents

\subsection{TBG continuum model} \label{methods: tbg_continuum_model}

Let us review the continuum model used to capture the non-interacting bandstructure of TBG. The parametrisation of the non-interacting Hamiltonian and the mean-field treatment of the Coulomb interactions follows the procedure outlined in Ref.~\cite{NadjPerge}, which we reproduce here for completeness.

For the non-interacting Hamiltonian, $\mathcal{H}_0$, we employ the continuum model introduced in Ref.~\citenum{MK18}.  As described in Eq.~\ref{eq:ham_cont} of the main text, the non-interacting Hamiltonian is given by
\begin{equation}
\mathcal{H}_0 = \sum_{\gamma = \{\xi,\sigma\}} \int_\Omega d^2\r~ \psi_\gamma^\dagger(\r)  \hat{H}^{(\xi,\sigma)} \psi_{\gamma}(\r),\quad
\hat{H}^{(\xi,\sigma)} = \begin{pmatrix} 
H_{\xi 1} & U_{\xi}^\dagger(\r) \\
U_{\xi}(\r) & H_{\xi 2} 
\end{pmatrix},\label{eq:app_ham_cont}
\end{equation}
where the explicit form of the elements of the operator $\hat{H}^{(\xi,\sigma)}$ are detailed below. The spinor, $\psi_\gamma$, is written in the basis of $(A_1, B_1, A_2, B_2)$ sites of the original two layers ($l=1,2$) and we use the shorthand notation, $\gamma\equiv\{\xi(=\pm1),\sigma(=\pm1)\}$, for the valley/spin degrees of freedom. The real space integration is over a moir\'e unit cell $\Omega$. The intralayer elements of the $\hat{H}^{(\xi,\sigma)}$ are
\begin{equation}
H_{\xi l}= -\hbar v \left[ R\left(\pm\theta/2\right)(\vec{k}-\vec{K}_{\xi}^{(l)})\right] \cdot \left(\xi\sigma_x,\sigma_y\right)
,
\label{eq:linear_equation}
\end{equation}
where $\vec{k}$ is a momentum in the BZ of the original graphene layers,  $R\left(\phi\right)$ is the $2\times 2$ two-dimensional counterclockwise rotation matrix that accounts for rotation of the BZs for the original graphene layers, and the signs $\pm$ in Eq.~\eqref{eq:linear_equation} correspond to the layers $l = 1$ and $2$, respectively. For all twist angles studied in the paper we use $\hbar v/a = 2.1354$ eV as the energy scale for the Hamiltonians $H_{\xi l}$. In Eq.\eqref{eq:linear_equation} above  the $\vec{k}\cdot \vec{p}$ expansion is taken around the two vectors $\vec{K}_{1}^{(l)}$, $\vec{K}_{-1}^{(l)}$, which denote the Dirac points $K$ and $K'$ of the two layers, respectively:
\begin{equation}
\vec{K}_{\xi}^{(1)} = -\xi \frac{4\pi}{3a} R\left(-\theta/2\right) \begin{pmatrix} 
1 \\
0
\end{pmatrix},\quad 
\vec{K}_{\xi}^{(2)} = -\xi \frac{4\pi}{3a} R\left(\theta/2\right) \begin{pmatrix} 
1 \\
0
\end{pmatrix}
\,
\end{equation}
Within the convention of Ref.~\cite{MK18} the moir{\'e} superlattice BZ is defined using two reciprocal lattice vectors
\begin{equation}
    \vec{G}_1^{M} = -\frac{2\pi}{\sqrt{3} L_M} \begin{pmatrix} 
1 \\
\sqrt{3} 
\end{pmatrix}
,\quad 
\vec{G}_2^{M} = \frac{4\pi}{\sqrt{3} L_M} \begin{pmatrix} 
1 \\
0 
\end{pmatrix}
\,,
\end{equation}
with $L_M = a/2\sin(\theta/2)$ being the moir{\'e} effective lattice period and $a=0.246$ nm corresponding to the lattice constant of graphene. We denote the reciprocal lattice vector length as $G_M = |\vec{G}_1^{M}| = |\vec{G}_2^M| = 4\pi / \sqrt{3} L_M$. The operator $U_{\xi}(\r)$ in Eq.\eqref{eq:app_ham_cont} encodes the interlayer hopping and is given by:
\begin{align}
\label{eq:app_moire_interlayer_coupling}
U = \begin{pmatrix}
u & u'\\
u' & u\\\end{pmatrix}&+\begin{pmatrix}
u & u'e^{-i2\pi \xi/3}\\
u'e^{i2\pi \xi/3} & u\\\end{pmatrix}e^{i\xi \vec{G}_{1}^M\cdot\vec{r}}+\begin{pmatrix}
u & u'e^{i2\pi \xi/3}\\
u'e^{-i2\pi \xi/3} & u\\\end{pmatrix}e^{i\xi \left(\vec{G}_{1}^M+\vec{G}_{2}^M\right)\cdot\vec{r}}\,
\end{align}
We treat the interlayer couplings $u$ and $u'$ as fitting parameters for the band structure according to the procedure introduced in Ref.~\cite{NadjPerge} and summarized below. To determine the energy bands and the eigenstates of both the non-interacting model and the one that includes mean-field corrections, we take the Bloch wavefunction ansatz for valley $\xi$ as
\begin{equation}
\Psi_{\xi,n,\vec{k}}^{j}(\vec{r})=\sum_{\vec{G}} C_{\xi,n,\vec{k}}^{j}(\vec{G})e^{i(\vec{k}+\vec{G})\cdot \vec{r}},\label{eq:bloch_ansatz}
\end{equation}
where $j=A_1$, $B_1$, $A_2$, $B_2$ labels the spinor components, $n$ is a band index, and $\vec{k}$ is the Bloch wave vector in the BZ of the original graphene layers. In the above ansatz the $\vec{G}$ sum runs over all possible integer combinations of the reciprocal lattice vectors $\vec{G}=m_1 \vec{G}_{1}^{M}+ m_2 \vec{G}_{2}^{M}$ with integer $m_1$ and $m_2$. In practice we constrain $-15 \leq m_1, m_2 \leq 15$ thus ensuring that the mean-field band-flattening and pairing calculations do not suffer from any non-interacting model cut-off effects.

\subsection{Mean-field treatment of interactions} \label{methods: tbg_mean_field}

In this section, we provide explicit forms for the mean-field Hartree and Fock potentials defined in the main text. The Hartree self-energy, $\mathcal{H}_H$ of Eq.\eqref{eq:Hartree}, when expressed in the basis used for the Bloch ansatz from Eq.\eqref{eq:bloch_ansatz}, takes the form
\begin{align}
\label{eq:app_hartree_self_energy}
    \left \langle \vec{k}+\vec{G},\gamma, i \left \lvert \mathcal{H}_H \right \rvert \vec{k}+\vec{G'}, \gamma' ,i' \right \rangle =
    \delta_{\gamma,\gamma'} \delta_{i,i'} \V_{\mrm{c}}(\vec{G}-\vec{G'}) \sum_{\gamma'',\vec{k}',\vec{G}'',i''}'\left(C_{\gamma'' \vec{k}'}^{i''}(\vec{G}'')\right)^* C_{\gamma'' \vec{k}'}^{i''}(\vec{G}''-\vec{G}+\vec{G}')\,.
\end{align}
In the above expression $\sum'$ denotes summation over occupied states for a given filling measured with respect to the CNP as explained in the main text (denoted $\left\langle \dots \right\rangle_H$). As explained in Refs.~\citenum{guineaElectrostaticEffectsBand2018,ceaElectronicBandStructure2019, ceaBandStructureInsulating2020}, since the Hartree potential is controlled primarily by the contribution of the first-star of reciprocal lattice vectors, it is sufficient to consider combinations of $\vec{G}$, $\vec{G'}$ satisfying $\vec{G}-\vec{G'}=m \vec{G}_{1}^M+n \vec{G}_{1}^M$ with $(m,n)=\{(\pm 1, 0), (0, \pm 1), (\pm 1, \pm 1)\}$. As argued in Ref.\cite{ceaElectronicBandStructure2019, ceaBandStructureInsulating2020}, going beyond the first-star of reciprocal vectors does alter results drastically. We note however that this is one explicit cut-off used in the calculation. As usual the $\vec{G}=\vec{G'}=0$ contribution is excluded since it is cancelled by the positive (jellium) ionic background.

Similarly, the Fock self-energy, $\mathcal{H}_F$, in Eq.~\eqref{eq:Fock} becomes
\begin{align}
    \label{eq:app_fock_self_energy}
    \left \langle \vec{k}+\vec{G},\gamma, i \left \lvert  \mathcal{H}_{F} \right \rvert \vec{k}+\vec{G'}, \gamma' ,i' \right \rangle = -\delta_{\gamma,\gamma'}\sum_{\vec{k}',\vec{G}''}'' \V_{\mrm{c}}(\vec{k}-\vec{p}+\vec{G'}-\vec{G}'') \left(C_{\gamma \vec{k}'}^{i'}(\vec{G}'')\right)^* C_{\gamma \vec{k}'}^{i}(\vec{G}''+\vec{G}-\vec{G}')\,,
\end{align}
where $\sum''$ denotes summation over occupied states, i.e. $\left\langle \dots \right\rangle_F$ of Eq.\eqref{eq:Fock} from the main text, in a manner explained in Sec. \ref{methods: parameters} below. Importantly, the self-energy explicitly depends on the crystal momentum $\vec{k}$ due to the non-local nature of the Fock potential. This dependence imposes significant numerical complexity for self-consistent calculations, unlike the Hartree form of Eq.~\eqref{eq:app_hartree_self_energy}, since the self-energy due to the Fock term has to be determined separately for each momentum $\vec{k}$. 

\subsection{Procedure for fitting TBG continuum model to scanning tunneling microscopy (STM) experiments} \label{methods: parameters}

Here we elaborate further on the modeling and analysis introduced in Sec. \ref{methods: tbg_continuum_model}, \ref{methods: tbg_mean_field}, specifically focusing on the relation to experiments. The specific fitting procedure follows closely the one adopted in Ref.\cite{NadjPerge} and is meant to be an accompanying reference. Our key objective here is to provide an experimentally inspired modelling of the qualitative behavior of the bandstructure and corresponding Bloch wavefunctions as a function of twist angle and filling, rather then provide a complete and full treatment of Hartree-Fock problem in TBG. We also note the large body of existing work in the literature that analyzes Hartree-Fock contributions (e.g.,  Refs.\citenum{ceaElectronicBandStructure2019,ceaBandStructureInsulating2020,bultinckGroundStateHidden2020,khalafChargedSkyrmionsTopological2020,xieNatureCorrelatedInsulator2020,xieWeakfieldHallResistivity2020}). Finally, as already emphasized in the main text, we do not address here the detailed nature of cascade states, and the presence or absence of Fock-induced gaps at integer fillings. 

The non-interacting continuum model as introduced in the Sec. \ref{methods: tbg_continuum_model} for a given twist angle has two free parameters - the interlayer couplings $u,u'$. Although their dependence on twist angle has been studied through ab-initio methods \cite{carrExactContinuumModel2019}, here we choose a simpler approach intended to highlight the important interaction-driven qualitative changes to the band structure. We assume that the two parameters $u$ and $u'$, corresponding to same-sublattice and opposite-sublattice interlayer 
tunneling, have fixed values for all twist angles. This approximation misses 
the subtle role relaxation physics plays on increasing the ratio of these parameters $\eta = u/u'$ as the twist angle is brought closer to the magic angle \cite{carrExactContinuumModel2019}. To fix $u$ and $u'$, we focus on the measurements at the largest available angle of $\theta=1.32^\circ$ in Ref.\cite{NadjPerge}, where the role of interactions is least important. By matching the measured Landau-level (LL) spectrum to that obtained numerically from this continuum model, we fix $u'=90$ meV and $u=0.4 u'$. As noted in the main text, as a result of this approximation scheme the magic-angle of the non-interacting model occurs at $\theta\approx 0.99^\circ$, which differs from the typical values $\theta\approx 1.1\degree$ quoted in the literature. At the same time however, the twist dependent ratio of $\eta = u/u'$ will not qualitatively alter our results. Specifically, the location of van Hove singularities in each band does not change with $\eta$ for values of $\eta \lesssim 0.8$ (See also discussion in Ref. \cite{2021arXiv210210504Q}) Combining this observation with the property that the peak of the pairing dome (within weak-coupling limit) is dictated by the location of the van Hove singularity, we argue that band-flattening corrections to pairing can be disentangled from any of those coming from varying $(u,u’)$. 

We now proceed to parametrize the strength of the dielectric screening $\epsilon$ in the Coulomb potential $\V_{\mrm{c}}(q)=2\pi e^2/\epsilon q$. Nominally the value set by the substrate, i.e. typical of an hBN encapsulated graphene is $\epsilon \approx 5$. This value however massively overestimates the role of Hartree and Fock processes, leading to band structures with large $\bar{\Gamma}$ point inversions that are not observed experimentally. To overcome this unwarranted behavior, earlier works \cite{guineaElectrostaticEffectsBand2018,ceaElectronicBandStructure2019,xieWeakfieldHallResistivity2020} use a range of values for $\epsilon$ ranging from $5$ 
to $66$. In the same spirit, we choose $\epsilon=15$ to quantitatively capture the following three 
experimental characteristics seen in the LL spectra of Ref.\cite{NadjPerge}: (i) the energy spacing between LLs arising from 
the $\bar{\Gamma}$ point band structure at $\theta=1.32^\circ$ (Fig. 1 in Ref.\cite{NadjPerge}), (ii) the energy spacing between the highest energy LL from the flat  band and the lowest energy LL from the dispersive bands at $\theta=1.32^\circ$ (Fig. 1 in Ref.\cite{NadjPerge}), and (iii) the critical angle at which largest energy LL from the flat-band joins the vHs (Fig. 2 in Ref.\cite{NadjPerge}). These criteria are met with a choice of $\epsilon = 15$, which we then keep constant for all values of $\theta$. With this parametrisation it was found in Ref.\cite{NadjPerge} that including Hartree-only correction adequately captures experimental observations at $\theta=1.32^\circ$, suggesting that the Fock term plays a subdominant role at least at this large twist angle.

As mentioned in the previous Sec. \ref{methods: tbg_mean_field}, the Fock potential is non-local and thus  carries a high computational cost, making a parameter sweep like that in Fig.~\ref{fig:results_T_C} prohibitive. Motivated by the physical intuition that the role of Fock is to oppose the Hartree potential, further substantiated by recent works \cite{xieWeakfieldHallResistivity2020}, we approximate Eq.~\eqref{eq:app_fock_self_energy} as
\begin{equation}
    \label{eq:app_approximated_fock_self_energy}
    \left \langle \vec{k}+\vec{G},\gamma, i \left \lvert \mathcal{H}_{F} \right \rvert \vec{k}+\vec{G'}, \gamma' ,i' \right \rangle \approx -\delta_{\gamma,\gamma'}  g(\theta)  \V_{\mrm{c}}(\vec{G}-\vec{G'}) \sum_{\vec{k}',\vec{G}''}'' \left(C_{\gamma \vec{k}'}^{i'}(\vec{G}'')\right)^* C_{\gamma \vec{k}'}^{i}(\vec{G}''+\vec{G}-\vec{G}')\,.
\end{equation}
In particular, we replace the non-local dependence with a constant, twist-angle dependent prefactor $g(\theta)$ with the characteristic energy scale of the Fock interaction being set by the Hartree potential $\V_{\mrm{c}}(\vec{G}-\vec{G'})$. We stress that despite similarity to a local approximation, e.g., $\V_{\mrm{c}}(\vec{r})\propto\delta(\vec{r})$, the form of Eq.~\eqref{eq:app_approximated_fock_self_energy} carries the additional dependence on reciprocal momenta $\vec{G},\vec{G}'$. In analogy to the Hartree potential, we limit the reciprocal momenta included in the self-energy to a difference $\vec{G}-\vec{G}'$ residing in the first star of reciprocal lattice vectors. Unsurprisingly, the Fock potential so constructed acts in opposition to the Hartree potential due to the overall minus sign above. We reiterate here that this is a phenomenological approximation, which nonetheless captures qualative behavior of the Fock term reported by other authors, e.g. Refs.~\cite{ceaElectronicBandStructure2019,ceaBandStructureInsulating2020, xieWeakfieldHallResistivity2020}. Namely it demonstrates how the Fock term can counteract band inversion stemming from the Hartree term; at charge neutrality point it also predicts an increase in the van Hove to van Hove separation and a broadening (splitting) of each flat-band's van Hove peak - both features seen in STM experiments \cite{Kerelsky2019,10.1038/s41567-019-0606-5, NadjPerge}. For a further discussion of this approach together with other possible mechanisms (e.g. strain) that can similarly prevent Hartree-driven band inversions we refer to Ref. \cite{NadjPerge}.

In Ref.\cite{NadjPerge} several forms for the prefactor of the Fock interaction $g(\theta)$ were considered. It was observed that a twist-angle-independent prefactor does not manage to capture the qualitative band structure behavior seen in experiments: A constant value that is too large 
prevents unphysical band inversion near the magic angle, yet it also prevents band flattening for $\nu=3,4$ seen in the experiment at intermediate angles that could be recovered  with a Hartree-only analysis;  A constant value that is unusually small allows for band flattening seen in experiments, but does not prevent unphysical band inversions near the magic angle. To remedy these issues in Ref.\cite{NadjPerge}, the following $g(\theta)$ dependence was chosen which we use in our analysis as well: we take $g(\theta)=0$ for $\theta \ge 1.14^\circ$ and $\theta \le 0.84^\circ$, whereas for $0.84^\circ \le \theta \le 1.14^\circ$ range we assume a triangular profile with a maximum of $g(\theta=0.99^\circ) = 1.875$.

A final element of the Fock term description is the specification of what bands are included in the summation $\sum''$ in Eqs.~\eqref{eq:app_fock_self_energy} and \eqref{eq:app_approximated_fock_self_energy} (or as schematically indicated with $\left\langle \dots \right\rangle_F$ in the main text). Several summation schemes are present in the literature with the key difference being whether only flat bands \cite{ceaElectronicBandStructure2019,ceaBandStructureInsulating2020} or also dispersive bands \cite{bultinckGroundStateHidden2020, xieNatureCorrelatedInsulator2020} are included. Qualitatively, at $\nu=4$ the Fock term arises predominantly from the flat bands; on the other hand, at $\nu=-4$ the dominant contribution arises from the dispersive valence bands. The contribution from the dispersive valence bands is also opposite in sign and effect to that of the flat bands as seen in Ref.~\citenum{xieWeakfieldHallResistivity2020}. We verified that this behavior holds within our approximation of Eq.~\eqref{eq:app_approximated_fock_self_energy}, but due to the local form of the Fock potential contribution from dispersive bands at $\nu=-4$, one finds gap opening near the $\bar{K},\bar{K}'$ points beyond what is experimentally plausible. To qualitatively capture the above sign trend in the Fock term while mitigating the preceding gap issue, we include in the summation $\sum''$ all flat band states both at $\nu=4$ and $\nu=-4$, but change the sign of the contribution at $\nu=-4$. For other fillings, motivated by Ref.~\citenum{xieWeakfieldHallResistivity2020}, we interpolate our solution as explained in the next paragraph. 

As the contribution of the Hartree and Fock terms to the self-energy depends linearly (in the absence of cascade and gap opening at the CNP) on filling\cite{ceaElectronicBandStructure2019, goodwinHartreeTheoryCalculations2020, xieWeakfieldHallResistivity2020}, for ease of computation we linearly interpolate between the solutions at $\nu=4$ and $\nu=-4$. We verified that the results do not qualitatively change if a self-consistent approach is used at every filling. To that end, the mean-field interacting Hartree-Fock Hamiltonians for filling $\nu$ are taken to be
\begin{align}
    \mathcal{H}_F(\nu) &= \frac{1}{2} \left[\left(\mathcal{H}_F(\nu=4)+\mathcal{H}_F(\nu=-4)\right)+\frac{\nu}{4}\left(\mathcal{H}_F(\nu=4)-\mathcal{H}_F(\nu=-4)\right)\right]\\
    \label{eq:app_final_self_energy_interpolated_fock}
    \mathcal{H}_H(\nu) &= \frac{1}{2} \left[\left(\mathcal{H}_H(\nu=4)+\mathcal{H}_H(\nu=-4)\right)+\frac{\nu}{4}\left(\mathcal{H}_H(\nu=4)-\mathcal{H}_H(\nu=-4)\right)\right]\,
\end{align}
where $\mathcal{H}_H(\nu)$, $\mathcal{H}_F(\nu)$ denotes Hartree or Fock correction for a filling $\nu$. For the bandstructures and calculations presented in this paper, we evaluate the solutions at $\nu=4$ and $\nu=-4$ self-consistently until convergence is reached. We set the convergence threshold for all self-consistent parameters as $0.1\%$ total relative error (difference between successive self-consistent steps). A grid of 441 $k$-points was used for the analysis of the self-consistent potentials (see next section for parametrisation of the SC calculation), where the convergence reached after a few  (typically in less than $5$) iterations. 

\subsection{Numerical solution of the linear gap equations} \label{app:numsol}

Here we provide additional technical details of the Eliashberg procedure we adopt in this paper. We note again that we have used the same framework as detailed in our earlier work Ref.\cite{CL20}. For a given twist angle and filling we start the calculation by pre-computing a 2D MBZ mesh of points, with their associated Bloch wavefunctions and energies. We note that because of the HF-corrections that enter at a mean-field level into this calculation, a different 2D mesh of points has to be precomputed for each filling. In the calculations we use a mesh of $10201$ MBZ points for each $\nu$ and $\theta$. To carry out an angle-average of the kernel from Eq.~\eqref{eq:Kernel}, we first fix a particular direction of vector $\vec{k}$ (upon verifying that the conclusions are not dependent on the specific direction) and then identify all $\vec{p}$ points that are of magnitude $p$ (within a resolution admitted by the mesh). We then estimate the angular average Eq.~\eqref{eq:Kernel} by averaging over these $\vec{p}$ points --- in practice, $\sim 50-100$ points are used for each pair of $k$ and $p$ momentum values.

To determine the critical temperature, we seek the temperature $T$ for which Eq.~\eqref{eq:sc_gap_equation} has an eigenvalue of unity. In practice, we carry out a bisection method search for a $T$ giving an eigenvalue within $\pm 0.01$ of unity. In the calculations we use a linearly spaced grid of $30$ $k$ points ranging from the center of the MBZ $\bar{\Gamma}$ to the $\bar{K}$ point. For the Matsubara grid, in analogy with the procedure of Ref.\cite{CL20} we employed an approximate scheme that consists of 10 first Matsubara frequencies followed by 20 linearly spaced frequencies starting from the 11th Matsubara frequency to the UV cutoff, where the cutoff is chosen to be $8W$ ($W$ is the flat-band bandwidth).

\begin{figure}[!h]
    \centering
    \includegraphics[width=\linewidth]{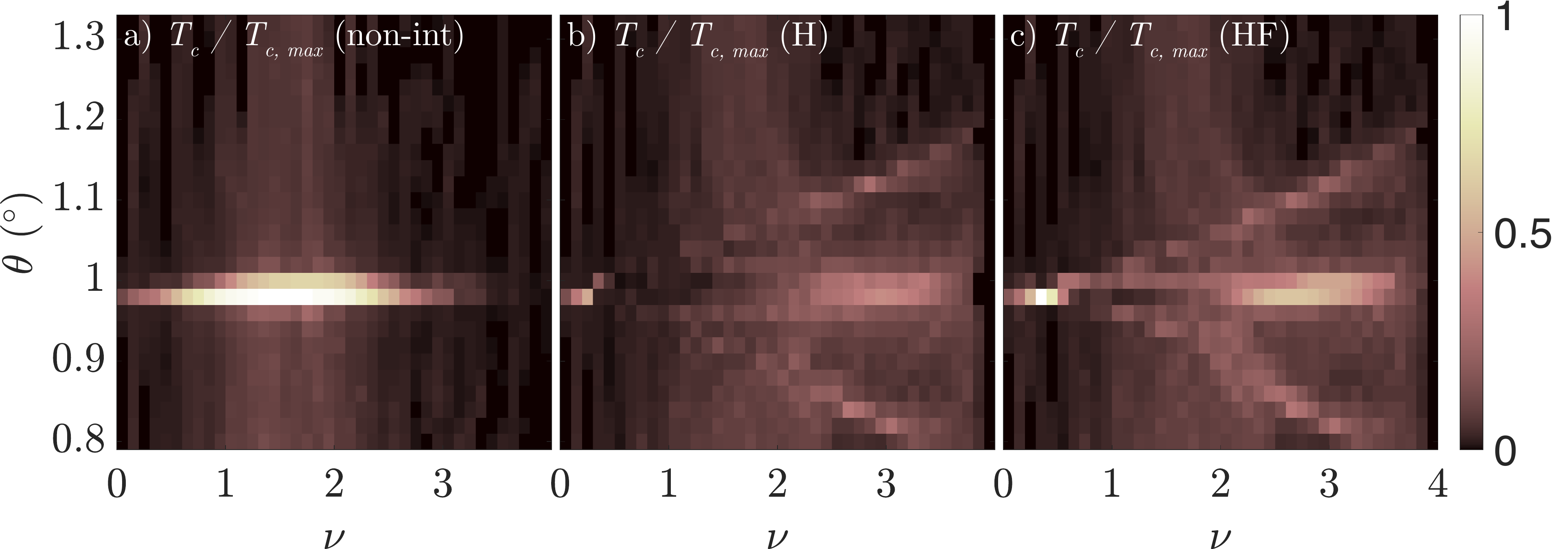}
    \caption{{\bf Role of Hartree and Fock corrections on the pairing temperature.} $T_c$ obtained from the linearized Eliashberg equation (Eq. \eqref{eq:sc_gap_equation}) as a function of $\nu$ and $\theta$ for (a) non-interacting, (b) Hartree-only corrected bandstructure, and (c) Hartree-Fock corrected bandstructure. The values of $T_c$ are normalized in (a), (b) and (c) relative to the highest pairing temperature $T_{C,max}$ in (a) (see discussion in the main text).}
    \label{fig:ex_fig_1}
\end{figure} 

\begin{figure}
    \centering
    \includegraphics[width=\linewidth]{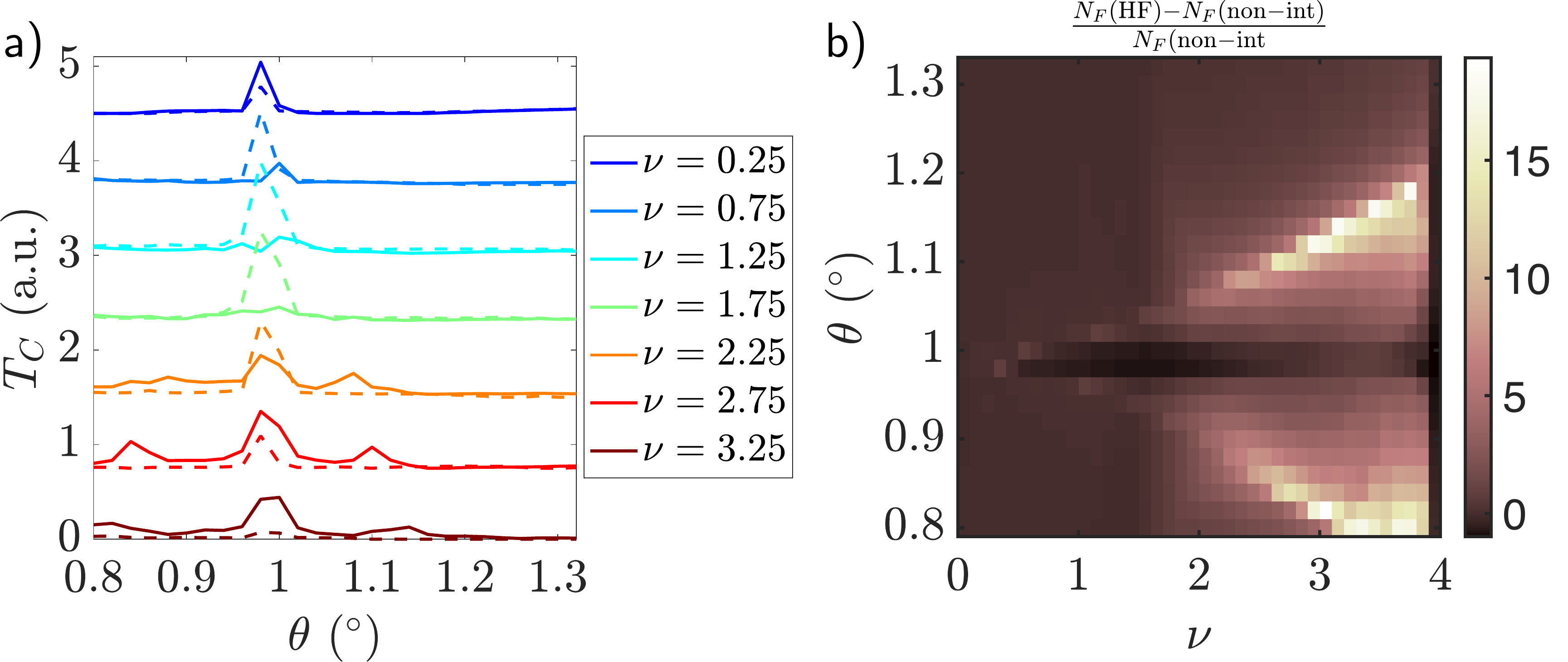}
    \caption{{\bf A comparison of the pairing tendencies in the non-interacting and HF modified system.} (a) constant filling linecuts of the Fig.\ref{fig:results_T_C}a,b for non-interacting (dashed) and HF-modified (solid) system. Curves are vertically separated for clarity. (b) Relative change of the Fermi surface density of states in the two models. Note how (a,b) demonstrated broadening of the twist angle range over which pairing tendencies can be seen.}
    \label{fig:ex_linecuts}
\end{figure}

\subsection{Comparison of Hartree vs. Hartree-Fock and role of electron-phonon umklapp} \label{methods:additional_results}

Here we present additional results for the calculation of $T_c$ as a function of filling and twist angle; we focus specifically on (i) a comparison of a non-interacting, Hartree-only and Hartree-Fock modified bandstructure on pairing tendencies, (ii) the role played by Umklapp processes, and (iii) the shape of the $T_c$ dome for negative fillings $\nu<0$.

In Supplemental Figure ~\ref{fig:ex_fig_1} we show the pairing dome profile as a function of twist angle and filling for three different ``extents'' of incorporating Coulomb interactions. Supplemental Figure \ref{fig:ex_fig_1}a shows a pairing profile for a non-interacting bandstructure and wavefunctions which, as discussed in the main text, is sharply peaked near the magic angle and falls precipitously away from it. As expected of a weak-coupling calculation, the dome is peaked at the location of the vHs of the non-interacting bandstructure that remains fixed for all twist angles. When Hartree-only (Supplemental Figure \ref{fig:ex_fig_1}b) and Hartree-Fock (Supplemental Figure \ref{fig:ex_fig_1}c) corrections are introduced into the SC calculation, we find the appearance of a qualitatively similar-looking three-prong feature stemming from band-flattening physics analysed in the  main text. The Hartree-only correction however presents an overall lower maximum $T_c$ near the magic angle as compared to the Hartree-Fock results. This simply is a consequence of the Hartree-only bandstructure featuring a large $\bar{\Gamma}$ point inversion that increases overall bandwidth at the magic angle beyond that of non-interacting or Hartree-Fock (See also Supplemental Figure \ref{fig:ex_linecuts} for further comparison of non-interacting and Hartree-Fock results.). In Supplemental Figure \ref{fig:ex_fig_1}, \ref{fig:ex_linecuts}, as in the main text, we normalize the pairing temperature $T_c$ by its maximum value for the non-interacting bandstructure of $T_{c, max}= 0.0294$ meV.

\begin{figure}
    \centering
    \includegraphics[width=\linewidth]{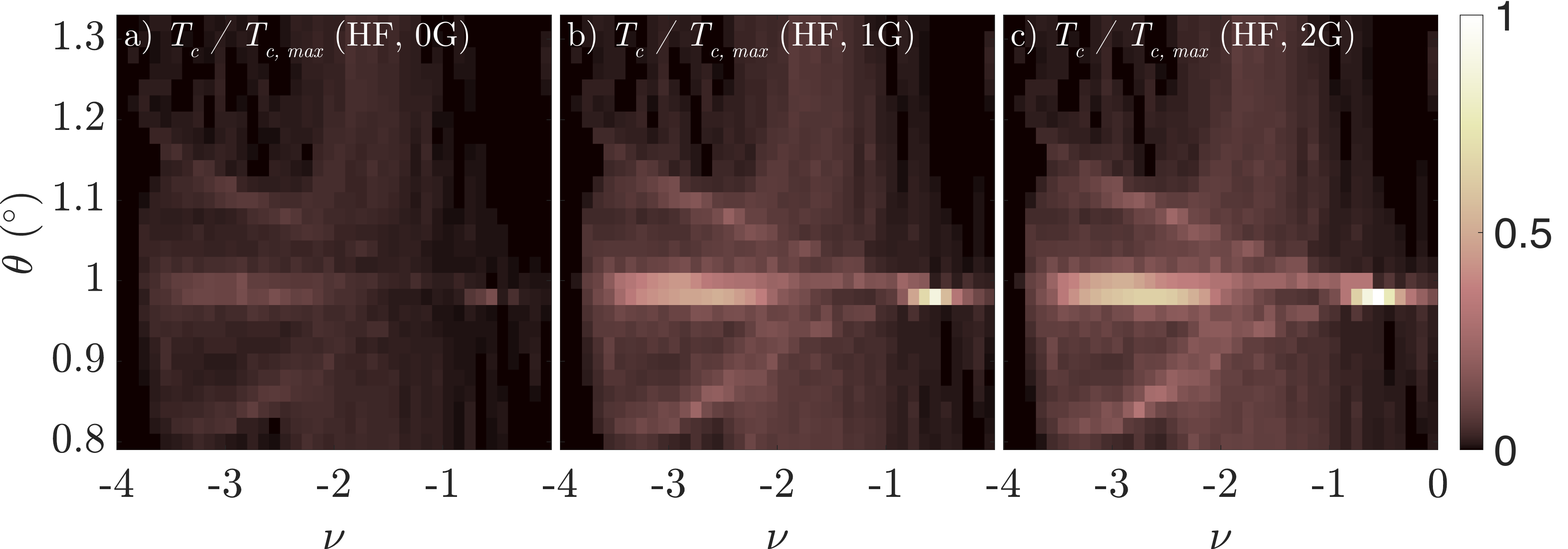}
    \caption{{\bf Contribution of phonon umklapp processes in enhancing pairing tendencies.} $T_c$ obtained from the linearized Eliashberg equation (Eq. \eqref{eq:sc_gap_equation}) as a function of $\nu$ and $\theta$ for (a) $0$G, (b) $1$G, and (c) $2$G phonon umklapp processes for a Hartree-Fock corrected bandstructure. The values of $T_c$ are normalized in (a), (b) and (c) relative to the highest pairing temperature $T_{C,max}=0.0294$ meV of Supplemental Figure~\ref{fig:ex_fig_1}a (see discussion in text).}
    \label{fig:ex_fig_2}
\end{figure}

In Supplemental Figure~\ref{fig:ex_fig_2} we study the effect of including umklapp phonon processes on the pairing temperature. As we argued in Ref.\cite{CL20}, in TBG due to the small size of the effective MBZ, scattering processes that involve an exchange of multiples of reciprocal lattice vectors (umklapp processes) are favourable. We denote a computation which incorporates an exchange of up to $m$ reciprocal lattice vectors $G_1^M$, $G_2^M$ as an ``$mG$" calculation. Due to the nature of the moir{\'e} potential that couples both graphene layers to each other, to construct a low energy effective theory it is necessary to use several graphene states that are coupled by these reciprocal lattice vectors $G$. The specific number of such states imposes a critical value of $m$ umklapp processes that can be exchanged above which there is no significant increase in pairing temperature - we refer interested reader to our earlier work Ref.\cite{CL20}. We stress that again the highest $m$ above which no further increase in pairing temperature (typically $mG=3G$) occurs is well below any cut-off imposed by the finite truncation of the continuum Hamiltonian (which would correspond to $mG=15G$ as specified in the Methods section of the main text. Indeed this physics finds representation in our results of Supplemental Figure~\ref{fig:ex_fig_2}a-c, where inclusion of higher number of umklapp processes enhances pairing temperature. We also note in passing that this effect becomes less pronounced the larger is the twist angle as the MBZ size increases, thereby suppressing umklapp processes. Finally, note that Supplemental Figure~\ref{fig:ex_fig_2}a-c is evaluated for negative fillings to provide an explicit demonstration of behavior that is qualitatively similar to $\nu>0$ (i.e. we do not find discernible qualitative differences between results of Fig.\ref{fig:results_T_C}b and Supplemental Figure~\ref{fig:ex_fig_1}c beyond some small quantitative differences stemming from particle-hole asymmetry that is present in the TBG Hamiltonian).

\end{document}